\documentclass[12pt,final]{article}%manuscript
\usepackage{graphicx}% Include figure files
\usepackage{epsfig}
\usepackage{dcolumn}% Align table columns on decimal point
\usepackage{bm}% bold math
\usepackage{achemso}
\usepackage{amssymb,amsmath}
\usepackage{gensymb}
\usepackage{multirow}
\usepackage{booktabs}
\makeatletter
\def\@biblabel#1{(#1)}
\makeatother
\usepackage{setspace}

\usepackage{booktabs,rotating}
\usepackage{fancyhdr}
\usepackage{booktabs,caption}
\usepackage[flushleft]{threeparttable}
\usepackage{enumerate,subfigure,tabularx,longtable}
\usepackage{longtable}
\usepackage[utf8x]{inputenc}
\UseRawInputEncoding
\usepackage{gensymb}
\usepackage {threeparttable} 

\usepackage{natbib}
\setkeys{acs}{articletitle = true}

\newcommand{\comment}[1]{}

\def\gsim {\mbox{\hbox{ \lower-.6ex\hbox{$>$}
\kern-1.12em \lower.5ex\hbox{$\sim$}\kern+.35em}}}
\def\lsim {\mbox{\hbox{ \lower-.6ex\hbox{$<$}
\kern-1.12em \lower.5ex\hbox{$\sim$}\kern+.35em}}}
\makeatletter
\let\@fnsymbol\@arabic
\makeatother

\begin{document}
% You should use BibTeX and revtex.bst for references

% marks overfull lines with blackboxes
%\draft
% Use the \preprint command to place your local institutional report
% number on the title page in preprint mode.
% Multiple \preprint commands are allowed.
%\preprint{3}

%Title of paper
\title{\vspace{-3.0cm}Molecular Perspectives  of \\
Interfacial Properties in the Water+Hydrogen \\
 System	in Contact with Silica or Kerogen
}

%\comment{ 
\author{Yafan Yang\thanks{To whom correspondence should be addressed, email: yafan.yang@cumt.edu.cn} $^{,\S}$, 
	Arun Kumar Narayanan Nair$^\dag$,\\
	Weiwei Zhu$^\ddag$,
	Shuxun Sang\thanks{To whom correspondence should be addressed, email: shxsang@cumt.edu.cn} $^{,\sharp,\natural}$,  %,\flat
	and Shuyu Sun\thanks{To whom correspondence should be addressed, email: shuyu.sun@kaust.edu.sa} $^{,\dag}$\\
	$^{\S}$ State Key Laboratory for Geomechanics and Deep \\
    Underground Engineering, China University of Mining \\
    and Technology,	Xuzhou, 221116, Jiangsu, China. \\
	$^\dag$ Computational Transport Phenomena Laboratory, Physical \\
    Science and Engineering Division, King Abdullah University of \\
	 Science and Technology, Thuwal, 23955-6900, Saudi Arabia. \\
	$^\ddag$ Department of Engineering Mechanics,\\
	 Tsinghua University, Beijing, 100084, China. \\
	$\sharp$ School of Resource and Earth Science, China University of \\
     Mining and Technology,	Xuzhou, 221116, Jiangsu, China. \\
	$\natural$ Key Laboratory of Coalbed Methane Resources and Reservoir   \\
	Formation Process, Ministry of Education, China University of\\
	 Mining and Technology,	Xuzhou, 221116, Jiangsu, China. \\	
}
%}

\date{\today}
\maketitle
\newpage
%\linenumbers
\begin{abstract}

Interfacial behaviours in multiphase systems containing H$_2$ are crucial to underground H$_2$ storage but are not well understood.	
Molecular dynamics simulations were conducted to study interfacial properties of the H$_2$O+H$_2$ and H$_2$O+H$_2$+silica/kerogen systems over a wide range of temperatures (298 - 523 K) and pressures (1 - 160 MPa).
The combination of the H$_2$ model with the INTERFACE force field and TIP4P/2005 H$_2$O model can accurately predict the interfacial tensions (IFTs) from the experiment.
The IFTs from simulations are also in good agreement with those from the density gradient theory coupled to the PC-SAFT equation of state.
Generally, the IFTs decrease with pressure and temperature. However, at relatively high temperatures and pressures, the IFTs increase with pressure. 
The opposite pressure effect on IFTs can be explained by the inversion of the sign of the relative adsorption of H$_2$. 
The enrichment of H$_2$ in the interfacial regions was observed in density profiles.
%%%
Meanwhile, the behaviours of contact angles (CAs) in the H$_2$O+H$_2$+silica system are noticeably different from those in the H$_2$O+H$_2$+kerogen system.
The H$_2$O CAs for the H$_2$O+H$_2$+silica and H$_2$O+H$_2$+kerogen systems increase with pressure and decrease with temperature. However, the effect of temperature and pressure on these CAs is less pronounced for the H$_2$O+H$_2$+silica system at low temperatures. 
The behaviours of CAs were understood based on the variations of IFTs in the H$_2$O+H$_2$ system (fluid-fluid interaction) and adhesion tensions (fluid-solid interaction).
Furthermore, the analysis of the atomic density profiles shows that the presence of H$_2$ in between the H$_2$O droplet and the silica/kerogen surface is almost negligible. 
Nevertheless, the adsorption of H$_2$O on the silica surface outside the H$_2$O droplet is strong, while less H$_2$O adsorption is seen on the kerogen surface. 

\end{abstract}


\newpage
\section{Introduction}

With the rapidly growing demand for the decarbonized energy supply, hydrogen as a clean fuel has drawn great attention since it can be potentially used to progressively replace fossil fuels.\cite{zohuri2019hydrogen,zivar2021underground}
Although the energy per mass of H$_2$ is large, the massive H$_2$ storage has been considered a serious problem mainly because of the low energy per volume of H$_2$.
Geological sites including salt caverns, deep saline aquifers, basaltic formations, coal seams, and depleted oil/gas reservoirs have been suggested for underground hydrogen storage.\cite{flesch2018hydrogen,tarkowski2019underground,muhammed2022review}
Hitherto, only salt caverns have been operated to store industrial scale of H$_2$.\cite{tarkowski2019underground,zivar2021underground}
Considering that salt caverns are not worldwide common, investigations on other types of storage sites are of great significance.
Structural, residual, adsorption, dissolution, and mineral trapping are typical mechanisms for trapping gas in geological formations.\cite{smit2014introduction,pan2021underground,iglauer2021hydrogen} 
Among them, structural and residual trapping are considered the most important mechanisms, where capillary forces are critical as it determines the capacity and stability of the gas storage.\cite{iglauer2022optimum,ide2007storage}

There are many experimental studies\cite{chow2018interfacial,chow2020erratum,hosseini2022h2,slowinski1957effect,massoudi1974effect,iglauer2021hydrogen,ali2021influence,sedev2022contact,higgs2022situ,esfandyari2022experimental,hosseini2022capillary} focusing on the interfacial properties of multiphase systems with H$_2$ that dictates the capillary force in geological formations.
For instance, the interfacial tensions (IFTs) of the H$_2$O+H$_2$ system decrease with pressure (1 - 40 MPa) and temperature (298 - 523 K), and the reduction of IFT with pressure is smaller at higher temperatures. \cite{chow2018interfacial,chow2020erratum,hosseini2022h2,slowinski1957effect,massoudi1974effect}
It has been shown that the contact angles (CAs) in the brine+H$_2$+quartz system fall in the range from 0$\degree$ to 50$\degree$.\cite{iglauer2021hydrogen}
Increasing pressure increases the CAs in the brine+H$_2$+silicate systems.\cite{iglauer2021hydrogen,ali2021influence}
The CAs of brine/H$_2$ on mica decrease with increasing temperature, while an opposite temperature effect on CAs on quartz was reported.\cite{iglauer2021hydrogen,ali2021influence} 
Meanwhile, the CAs in the brine+H$_2$+bituminous coal system increase with pressure at 298 K, and the pressure effect is moderate at 323 and 343 K.\cite{sedev2022contact}

Molecular simulations have been applied to understand the interfacial behaviours in the gas+water\cite{garrido2019physical,miqueu2011simultaneous,yang2017molecular,yang2019effect,yang2022interfacial} and gas+water+solid system.\cite{yang2022interfacial,chen2016pressure,jagadisan2022molecular,zhou2022mechanisms,ho2021molecular,yang2022interfacialJCP} 
The IFTs from molecular dynamics (MD) simulations are generally in good agreement with experimental data and density gradient theory (DGT) predictions.\cite{garrido2019physical,miqueu2011simultaneous,yang2019effect,yang2022interfacial}
The IFTs in the H$_2$O+CH$_4$ and H$_2$O+CO$_2$ system decrease with temperature,\cite{garrido2019physical,miqueu2011simultaneous,yang2017molecular,yang2019effect} and decrease first with pressure and then increase with pressure as  pressure increases.\cite{garrido2019physical,miqueu2011simultaneous}
The opposite pressure effect on IFT was attributed to the inversion of the sign of relative adsorptions of gas obtained from component density distributions.
The simulated CAs in the H$_2$O+CO$_2$+silica system increase as pressure increases and temperature decreases.\cite{chen2016pressure,yang2022interfacialJCP}
Similar temperature effects on CAs were reported in the H$_2$O+CH$_4$+kerogen system.\cite{jagadisan2022molecular}
Interestingly, the CAs from MD simulation of the H$_2$O+CO$_2$+kerogen system increases from 60 (H$_2$O-wet) to 180$\degree$ (CO$_2$-wet) when pressure increases from 0 to 44 MPa.\cite{zhou2022mechanisms} 
Moreover, the thickness of the CO$_2$ film between the H$_2$O droplet and the kerogen surface increases with pressure.
Nevertheless, molecular-level understanding of the multiphase systems with H$_2$ is lacking.

In this article, MD simulations were conducted to study interfacial properties of the H$_2$O+H$_2$ and H$_2$O+H$_2$+solid systems over a wide range of temperatures (298 - 523 K) and pressures (1 - 160 MPa).
Silica and kerogen were selected to represent the solid phase as they are abundantly found in geological formations.\cite{iglauer2021hydrogen,ali2021influence,sedev2022contact}
The simulation results of the fluid system were complemented by DGT calculations with the Perturbed-chain Statistical Associating Fluid Theory (PC-SAFT) equation of state (EoS).
Details regarding the calculation and analysis of the interfacial tension, relative adsorption, enrichment, solubility, contact angle, adhesion tension, density distribution, and capillary pressure can be found in the following sections.

\section{Methodology}
\subsection{Simulation details}
MD simulations were conducted using the LAMMPS\cite{plimpton1995fast} package. 
Our simulation system include H$_2$O, H$_2$, silica, or kerogen (see Fig.\ref{fig:z1_vmd}).
The interactions between $i$ and $j$ molecular sites of different molecules are treated according to a pairwise additive Lennard-Jones (LJ) 12-6 function: \cite{frenkel2001understanding}
\begin{equation}
	\label{eq:LJ}
	U_\mathrm{LJ}(r_{ij}) = 4 \varepsilon _{ij}\left[ \left(\frac{{\displaystyle
			\sigma_{ij}}}{{\displaystyle r_{ij}}}\right)^{12} -
	\left(\frac{{\displaystyle
			\sigma_{ij}}}{{\displaystyle r_{ij}}}\right)^{6} \right],
\end{equation}
where $r_{ij}$ is the distance between the centers of $i$ and $j$ sites.
The parameter $\varepsilon_{ij}$ controls the strength of the short-range interactions, and the LJ diameter $\sigma_{ij}$ is used to set the length scale. 
The LJ parameters $\sigma_{ij}$ and $\varepsilon _{ij}$ are deduced from the Lorentz-Berthelot combining rules:\cite{frenkel2001understanding}
\begin{equation}
	\label{eq:LBLJsigma}
	\sigma_{ij} =\frac{\sigma_{i}+ \sigma_{j}}{2},
\end{equation}
\begin{equation}
	\label{eq:LBLJepsilon}
	\varepsilon_{ij} =\sqrt{ \varepsilon_{i}\varepsilon_{j}},
\end{equation} 
The charged sites are interacting with each other via the Coulomb potential:
\begin{equation}
	\label{eq:U_COUL}
	U_\mathrm{Coul}(r_{ij}) =  \,
	\frac{q_iq_j}{4 \pi \varepsilon _0 r_{ij}}\,
\end{equation}
where $q_i$ and $q_j$ are the partial charges of the sites $i$ and $j$, respectively,
and $\varepsilon _0$ is the dielectric permittivity of vacuum.

The LJ parameters and charges of all the fluid molecules used in this study are presented in Table 1. The TIP4P/2005 H$_2$O model\cite{abascal2005general} was used because it can accurately predict the surface tension of H$_2$O\cite{ghoufi2019calculation}. 
Three different hydrogen models were selected to test their performance for evaluating interfacial properties of the H$_2$O+H$_2$ system. 
The first model is from the INTERFACE force field (IFF)\cite{wang2021accurate}. The IFF model includes two LJ spheres centered on the hydrogen nuclei. The IFF H$_2$ model is flexible and the harmonic bond model is used to evaluate the bonding energy:
\begin{equation}
	U_{Bond}= K_{bond} (r_{HH}-r_{0,HH})^2,
\end{equation}
where $K_{bond}$ is the force constant (350 Kcal/mol/\AA$^2$), $r_{HH}$ is the bond length, and $r_{0,HH}$ is the equilibrium bond length of H$_2$ molecule (0.7414 \AA). The second H$_2$ model includes a unique LJ sphere taken from Hirschfelder et al.\cite{hirsehfelder1954molecular}. The third model includes an LJ sphere and a quadruple moment described with three charged particles which properly reproduce the experimental gas-phase quadrupole moment of H$_2$ according to Alavi et al.\cite{alavi2005molecular}

%% solid silica and kerogen
Silica and kerogen were used to represent the solid phase for the wettability study. Silica (SiO$_2$) with Q$^3$/Q$^4$ surface environments was described using the IFF model.\cite{emami2014force} 
The silanol (-OH) density of this silica surfaces was 2.4 per nm$^{2}$ and these surfaces are assumed to be nonionized.\cite{emami2014force} 
It has been shown that the water CAs on silica surfaces varies with different density of silanol groups.\cite{chen2015water} In real geological formations, the true surface of quartz should be a combination of possible function groups (Q$^2$,Q$^3$,Q$^4$,SiO$^-$, et al.).\cite{chen2015water} To simplify the problem, we focus on one type of surface with the silanol density in between Q$^3$ and Q$^4$ surfaces.  
The entire silica framework, except the H-atoms, was considered to be rigid during simulations. 

The type II-D kerogen molecular model developed by Ungerer et al.\cite{ungerer2015molecular} was chosen to model mature kerogens. The chemical formula of the exemplified kerogen is C$_{175}$H$_{102}$O$_9$N$_4$S$_2$.
The aromaticity of this kerogen is around 79\%, and H/C and O/C ratios are 0.58 and 0.051 respectively.\cite{ungerer2015molecular} 
It is important to note that the type and the thermal maturity of the kerogen have significant effects on water CAs.\cite{jagadisan2022molecular} Here, the overmature stage of type II kerogen was selected as it is associated with unconventional gas reserves such as the Barnett shale.\cite{ungerer2015molecular}
The force field parameters for the kerogen macromolecule were taken from the consistent valence force field (CVFF).\cite{dauber1988structure} For the construction of the kerogen surface, we followed the protocol described in Jagadisan and Heidari.\cite{jagadisan2022molecular} Briefly, 36 kerogen molecules were first randomly placed in a simulation box. Then two LJ walls were inserted on both sides of the simulation box. One of the walls was rigid, while the other one was allowed to move. The kerogen molecules were then compressed together by exerting an external force on the moving wall. The final density of the kerogen plate is around 1.3 g/cm$^3$, which falls within the range of experimental data for type II kerogen (1.18-1.35 g/cm$^3$).\cite{okiongbo2005changes} All kerogen atoms are flexible during simulations.

%%%%%%%% figure S1
Fig. \ref{fig:z1_vmd}(a) presents the equilibrium snapshot of the H$_2$O+H$_{2}$ two-phase systems. We employed 2048 H$_2$O and up to 1400 H$_2$ molecules for this system.
The box sizes in the x- and y-direction were fixed to be 36 {\AA}, which were large enough to remove the finite-size effects.\cite{yang2017molecular,yang2022interfacial} Three-dimensional periodic boundary conditions were implemented. 
The equilibrium box length in the z-direction ($L_z$) is 3-7 times larger than the lateral cell length, depending on the temperature and pressure conditions. The velocity Verlet algorithm was employed to integrate the coupled Newton's equations. The $NP_zT$ (constant number of molecules, pressure in z-direction, and temperature) equilibration and $NVT$ (constant number of molecules, volume, and temperature) production runs were 5 and 5 ns, respectively. The Nos\'e-Hoover thermostat with a relaxation time of 100 fs and the Nos\'e-Hoover barostat with a relaxation time of 1000 fs were applied to control the temperature and pressure, respectively. 

The IFT in the two-phase system is evaluated from the pressure tensor of the simulation box according to the Kirkwood and Buff approach\cite{kirkwood1949statistical}:
\begin{equation}
	\gamma =  \frac{1}{2}L_z \Big [P_{zz}-\frac{1}{2}(P_{xx}+P_{yy})\Big ],
\end{equation}
where $P_{xx}$, $P_{yy}$, and $P_{zz}$ denote the three diagonal components of the pressure tensor and $L_z$ is the simulation box length in the $z$-direction. Three independent trajectories were generated with different initial conditions for evaluating the error bars. It has been shown that long-range interactions have a significant influence on the IFT.\cite{ghoufi2019calculation,stephan2020influence} 
In Fig. S1, we compared the surface tensions of H$_2$O from simulations with long-range LJ interaction calculated by the PPPM method and those from simulations with truncated LJ interaction with a cutoff of 15 \AA. Here, the long-range electrostatic interactions were considered in both cases. It can be seen that the simulation results considering long-range LJ interactions agree better with experiment data, especially at low temperatures. Hence, in this study, the long-range part of LJ dispersion and electrostatic interactions were evaluated by the PPPM method with a precision of 10$^{-4}$. And the cutoffs for Lennard-Jones and electrostatic interactions were 9.5 {\AA}, respectively. 

We employed around 2000 H$_2$O and up to 12000 H$_2$ molecules for studying the H$_2$O+H$_2$+silica and H$_2$O+H$_2$+kerogen systems (see Figs. \ref{fig:z1_vmd}(b) and (c)). The line tension effect is insignificant in our simulations with the use of a cylindrical H$_2$O droplet. \cite{tenney2014molecular,chen2015water}
The dimensions of the simulation cell were 194.9 {\AA} $\times$ 34.3 {\AA} $\times$ 230.0 {\AA} (thicknesses of silica and kerogen plate were approximately 26 and 17 {\AA}, respectively). A bounding piston (with short-ranged interactions) that can be moved in the $z$-direction is used to control the bulk pressure.\cite{javanbakht2015molecular} 
The $NPT$ equilibration and $NPT$ production runs were 6 and 12 ns, respectively. The contact angles were determined using polynomial fits to the density profiles of the water droplets. \cite{smirnov2020structure,tenney2014molecular,chen2015water} The error bars of contact angles were computed based on the standard deviation of averages of 4 blocks with the block length of 3 ns. 

%%%%%%

\subsection{Theoretical details}

PC-SAFT EoS was applied to estimate the bulk properties of the fluid mixture. This EoS can be expressed via the compressibility factor $Z$~\cite{gross2001perturbed,gross2002application}:
\begin{equation}
	\label{eq:Z}
	Z = 1+Z^{\rm hc}+Z^{\rm disp}+Z^{\rm assoc},
\end{equation}
where $Z^{\rm hc}$ is the hard-chain term, $Z^{\rm disp}$ is the dispersive part, and $Z^{\rm assoc}$ represents the contribution due to association.
$Z$ is a function of the segment number $m_i$, the segment diameter $\sigma _i$, and the segment energy parameter $\epsilon_i$. 
The parameters for a pair of unlike segments were estimated using the Lorentz-Berthelot combining rules~\cite{gross2001perturbed}:
\begin{equation}
	\sigma _{ij} = \frac{1}{2} (\sigma _i+\sigma_j),
\end{equation}
\begin{equation}
	\epsilon_{ij} = \sqrt{\epsilon_i\epsilon_j}(1-k_{ij}),
\end{equation}
where $k_{ij}$ is the binary interaction parameter. The EoS parameters for pure component were taken from previous studies,\cite{diamantonis2011evaluation,alanazi2022evaluation} and in the absence of literature data, the $k_{ij}$ for H$_2$O-H$_2$ pair was fitted to the experimental solubility data.\cite{wiebe1934solubility} All parameters are given in Table S1 and S2.

PC-SAFT EoS was coupled with the DGT for the estimation of interfacial properties. In DGT, for a planar interface of area $A$, the Helmholtz free energy is given as~\cite{davis1982stress,davis1996statistical}:
\begin{equation}
	\displaystyle F = A \int_{-\infty}^{+\infty} \Big[f_0(n)+ \frac{1}{2}\sum_{i}^{}\sum_{j}^{} c_{ij}\cfrac{d n_i}{dz} \cfrac{d n_j}{dz} \Big] dz,
\end{equation}
where $f_0$ denotes the Helmholtz free energy density of the homogeneous fluid at the local density $n$, $d n_i/dz$ represents the local density gradient of the $i$th component. The cross influence parameter $c_{ij}$ is defined as~\cite{cornelisse1993application,miqueu2011simultaneous}:
\begin{equation}
	c_{ij} = (1-\beta_{ij}) \sqrt{c_{ii} c_{jj}},
\end{equation}
where $c_{ii}$ and $c_{jj}$ represent the pure component influence parameters, and $\beta_{ij}$ denote the binary interaction coefficient. These parameters were taken either from previous study\cite{mairhofer2018modeling} or fitted to the experiment data,\cite{linstrom2001nist} which are provided in Tables S3 and S4.

In equilibrium, the density profiles across the interface were evaluated through the minimization of the free energy by solving the corresponding Euler-Lagrange \mbox{equation~\cite{davis1982stress,davis1996statistical}}:
\begin{equation}
	\displaystyle \sum_{j}^{} c_{ij} \cfrac{d^2 n_j}{dz^2} = \mu_i (n_1 (z), ..., n_{N_c} (z)) - \mu_i^0\ {\rm for} \ i,j = 1,..,N_c,
\end{equation}
where $\mu_i^0$ is the chemical potential of the $i$th component in the bulk phase ($\mu_i^0 \equiv (\frac{\partial f_0}{\partial n_i})_{T,V,n_j}$), $\mu_i$ represents the chemical potential of the $i$th component and $N_c$ denotes the total number of components.  
The nonlinear equations were discretized by a finite difference method and solved by the Newton-Raphson iteration with the in-house code. 
A total of 200 equidistant grid points were used.
The linear density profiles were taken as the initial guess.
The interfacial thickness $l$ is initially assumed to be 10 {\AA} and then gradually
increased until the convergence is reached for the IFT value.\cite{davis1982stress,davis1996statistical}
The boundary conditions were obtained from flash calculations\cite{rachford1952procedure,pan2019multiphase}:
\begin{equation}
	\begin{array}{l}
		n_i = n_i^I\ {\rm at} \ z = 0,\\
		n_i = n_i^{II}\ {\rm at} \ z = l,
	\end{array}
\end{equation}
where $n_i^I$ and $n_i^{II}$ represent the bulk densities of the coexisting phases. 
When the equilibrium density profiles were available, the interfacial tension ($\gamma$) was estimated as follows~\cite{davis1982stress,davis1996statistical}:
\begin{equation}
	\displaystyle \gamma = \int_{-\infty}^{+\infty}  \sum_{i}^{} \sum_{j}^{} c_{ij} \cfrac{d n_i}{dz} \cfrac{dn_j}{dz} dz.
\end{equation}

\section{Results and Discussion}

\subsection{Interfacial properties in the H$_2$O+H$_{2}$ system}

% fig 2, fig s2
In order to evaluate the performance of different H$_2$ models in the H$_2$O+H$_2$ system, we calculated the IFTs at different temperatures and pressures. The simulation results were compared with experimental data in Fig. \ref{fig:z2}. It can be seen that the simulation data are close to experimental data, although simulation underestimates the IFT by 1.7-7.4 mN/m. High temperature decreases the IFT in both simulation and experiment. Consistent with the experimental trend, the IFTs of systems with the IFF and Hirschfelder H$_2$ model decrease as pressure increases. However, the most sophisticated Alavi model fails to describe the correct pressure effect on IFT. The averaged absolute deviations of the IFF and Hirschfelder models are 5.5 and 6.4 \%, respectively. Hence, the IFF H$_2$ model is selected for this study. In addition, the IFF model and the PC-SAFT EoS parameters of H$_2$ can accurately predict the experimental H$_2$ density\cite{linstrom2001nist} as displayed in Fig. S2.

% fig 3(a)
Fig. \ref{fig:z3}(a) shows the IFTs from the MD simulation and DGT for the H$_2$O+H$_{2}$ two-phase system under 298-523 K and 1-160 MPa in comparison to the experimental data.\cite{chow2018interfacial,chow2020erratum} The simulation and DGT data agree well with the experiment. Other experimental data\cite{hosseini2022h2,slowinski1957effect,massoudi1974effect} are in good agreement with the data here and not included for clarity.
The calculated IFTs fall in the range of 25 to 72 mN/m under the selected conditions.
Consistent with the experiment, the IFTs from MD and DGT decrease with the temperature at all pressures. 
Similar temperature dependence on IFT has been reported in the H$_2$O+CO$_2$\cite{chalbaud2009interfacial}, the H$_2$O+CH$_4$\cite{wiegand1994interfacial}, and the H$_2$O+N$_2$\cite{wiegand1994interfacial} systems. 

Although the effect of pressure on the IFT of the H$_2$O+H$_2$ system is generally moderate in contrast to the temperature effect, the pressure dependence of IFT is complicated depending on pressure and temperature. Consistent with the experiment, the IFTs from MD and DGT decrease with pressure under relatively low-pressure conditions. Notably, the decrease of IFT due to pressure increase is weaker at elevated temperatures. However, under relatively high pressure or high temperature conditions, the IFTs increase as pressure increases. For example, at 448 K, a minimum IFT is located at around 40 MPa based on DGT predictions. The corresponding IFT first decreases from 42.84 to 42.57 as pressure increases from 2 to 40 MPa. When pressure increases from 40 to 160 MPa, the IFT increases from 42.57 to 44.00 mN/m. Moreover, the transition point for the inversion of the pressure effect on IFT decreases as temperature increases.
At 523 K, the IFT from MD and DGT increases monotonically as pressure increases. Although there is no experimental data reporting the behaviours of IFT in the H$_2$O+H$_2$ system under the same pressure and temperature ranges, similar complex pressure dependence on IFT has been reported in the H$_2$O+CO$_2$\cite{garrido2019physical}, the H$_2$O+CH$_4$\cite{miqueu2011simultaneous,wiegand1994interfacial}, and the H$_2$O+N$_2$\cite{wiegand1994interfacial} systems. 
The IFTs in those binary systems are generally much smaller than the H$_2$O surface tensions especially at high pressures, while the IFTs in the H$_2$O+H$_2$ systems are close to the H$_2$O surface tensions.
Remarkably, at around 373 K, the transition point for the H$_2$O+CO$_2$\cite{garrido2019physical}, the H$_2$O+CH$_4$\cite{wiegand1994interfacial}, the H$_2$O+N$_2$\cite{wiegand1994interfacial}, and the H$_2$O+H$_2$ systems (from DGT in this work) are about 175, 140, 140, and 130 MPa, respectively.

% fig 4 %fig S3 
The behaviour of IFT is closely related to the density profiles. While it is difficult to measure the density profiles across the interface experimentally due to the thermal fluctuation\cite{stephan2020enrichment}, the density profiles can be obtained in MD and DGT\cite{stephan2020enrichment,garrido2019physical,miqueu2011simultaneous,yang2022interfacialatmosphere,yang2020bulk,yang2022interfacialJML1,yang2022studyJML2,yang2022interfacialJML3}. Fig. \ref{fig:z4} display density distributions of H$_2$O and H$_{2}$ from MD and DGT at various temperatures and pressures. Density profiles from MD are in line with those from DGT. The density profile of H$_2$O decreases monotonically from the aqueous phase to the H$_2$-rich phase, while a maximum H$_2$ density can be seen in the interfacial region. The enrichment is frequently used to quantify the non-monotonicity of the component density profiles:\cite{becker2016interfacial,stephan2023monotonicity} 
\begin{equation}
	E_i =  \cfrac{max(n_i(z))}{max(n_i^{I},n_i^{II})},
\end{equation}
where $n_i(z)$ is the number density profile of component $i$ across the interface, and $n_i^{I}$ and $n_i^{II}$ indicate the density of component $i$ in bulk phases $I$ and $II$, respectively. 
The calculated $E_{H_2}$ from MD for Figs. \ref{fig:z4}(a)-(d) are 1.07, 1.02, 1.01, and 1.00, respectively, while the corresponding data from DGT are 1.42, 1.14, 1.10, and 1.05, respectively. The enrichments from DGT are larger than those from MD, and the H$_2$ enrichment decreases with temperature and pressure. Furthermore, the solubility of H$_2$O and H$_2$ calculated based on bulk densities were given in Fig. S3. It can be seen that the solubility of H$_2$ in the aqueous phase increases with pressure and temperature, while the solubility of H$_2$O in the H$_2$-rich phase decreases with pressure and increases with temperature. The solubilities shown here are also in reasonable agreement with previous experimental measurement.\cite{wiebe1934solubility,kling1991solubility,bartlett1927concentration,eller2022modeling} 

% fig 3(b)
The effect of component interfacial distributions on the IFT can be understood by analyzing the relative adsorption. The relationship between the IFT and the relative adsorption (also known as the surface excess) is given by the Gibbs adsorption equation\cite{radke2015gibbs,stephan2020enrichment,miqueu2011simultaneous}:
\begin{equation}
	\label{eq:Gibbs}
	- d\gamma = \sum\limits_{i} \Gamma_id\mu_i,
\end{equation}
where is $\Gamma_i$ the relative adsorption of the component $i$ relative to the reference component $j$. The relative adsorption can be calculated from component density distributions:\cite{telo1983structure,wadewitz1996density}
\begin{equation}
	\label{eq:SurExcess}
	\Gamma_{i,j} = -(n_{i}^{II} - n_{i}^{I}) \int_{-\infty}^{+\infty}  \left[ \cfrac{n_{j} (z) - n_{j}^{II}}{n_{j}^{II} - n_{j}^{I}} - \cfrac{n_{i} (z) - n_{i}^{II}}{n_{i}^{II} - n_{i}^{I}} \right] dz,\
\end{equation}
where $I$ denotes the $i$-rich phase and $II$ represents the $j$-rich phase. 
Fig. \ref{fig:z3}(b) shows the relative adsorption of H$_2$ at different conditions. 
The relative adsorptions of H$_2$ from MD simulation are in qualitative agreement with those from DGT and experiment. 
Note that the relative adsorptions of H$_2$ from the expeirment are calculated based on Eq.\ref{eq:Gibbs} using experimental IFT values\cite{chow2020erratum,chow2018interfacial} and fugacities of H$_2$O+H$_2$ mixtures from PC-SAFT EoS.
The behaviour of H$_2$ relative adsorption at relatively low temperatures is similar to CH$_4$ relative adsorption in the H$_2$O+CH$_4$ two-phase system in the previous study.\cite{miqueu2011simultaneous}. For example, at 298 K, the H$_2$ relative adsorptions are all positive, and increase first with pressure and then decrease with pressure after reaching a maximum point at around 40 MPa. At low pressures, the increase of H$_2$ relative adsorption with pressure may come from the build-up of the sorption layer in the liquid-vapor interface as more H$_2$ is available in the gas phase under relatively high pressure. However, at high pressures, the increase of bulk density of H$_2$ is larger than the increase of H$_2$ sorption in the interfacial region. Hence, the relative adsorption of H$_2$ decreases with pressure.

High temperature decreases the relative adsorptions of H$_2$. This is because, under higher temperatures, accumulated H$_2$ in the interfacial region are more dispersed due to greater thermal fluctuations. The high temperature also changes the trend of the curves from a nonmonotonic shape with a maximum to a monotonically decreasing shape. 
For example, at 523 K, H$_2$ relative adsorptions are all negative and decrease as pressure increases. By analyzing the Gibbs adsorption equation, the complicated pressure effect of IFT at various temperatures mentioned above can be mostly explained by the switch of the sign of the H$_2$ relative adsorption given that the change of chemical potential of H$_2$ with pressure is positive.

\subsection{Interfacial properties in the H$_2$O+H$_{2}$+silica system}

The simulated CA of H$_2$O on the silica surface was
validated in our previous studies.\cite{yang2022interfacialJML1,yang2022interfacialJCP} 
Fig. \ref{fig:z5}(a) shows the contact angles of H$_2$O droplet from the MD simulation for the H$_2$O+H$_{2}$+silica three-phase system under 298-523 K and 1-160 MPa. We see that the simulated CAs fall in the range from 21.8$\degree$ to 49.0$\degree$. At 298 and 373 K, the effect of temperature and pressure on CA is little. However, at relatively high temperatures, the CA decreases with temperature and increases with pressure. 
For instance, the CAs at 40 MPa decrease from 48.0$\degree$ to 27.2$\degree$ as temperature increases from 373 to 523 K. And the CAs at 523 K increase from 21.8$\degree$ to 38.7$\degree$ as pressure increases from 5 to 160 MPa. 
Moreover, the reduction of CA with temperature is pronounced at low pressures while the increment of CA with pressure is strong at high temperatures. At low pressures, the CAs here are similar to the CAs in the H$_2$O+silica system as shown in Fig. S4 (also see ref.\cite{yang2022interfacialJML1} for results with SPC/E H$_2$O model).
Consistent trends were reported previously for the simulated CAs of the H$_2$O+CO$_2$+silica system under high temperature and pressure.\cite{chen2016pressure,yang2022interfacialJCP} 
The CAs in the brine+H$_2$+quartz system from experiment increases from 8$\degree$ to 28$\degree$ as pressure increases from 5 to 25 MPa at 298 K.\cite{iglauer2021hydrogen} The increment of CAs with pressure is consistent with our results from MD simulation. Nevertheless, our values are higher than those from experiment, which is likely because of the low density of the silanol group on the simulated silica surface.\cite{chen2015water}

The adhesion tension ($\gamma_{\rm SH} -\gamma_{\rm SW}$) describes the contribution of the solid-fluid interactions to the contact angle according to Young's equation: \cite{young1805iii,bartell1928degree}
\begin{equation}
	cos \theta=\cfrac{\gamma_{\rm SH} -\gamma_{\rm SW}}{\gamma_{\rm WH} },
	\label {eq:LY} 
\end{equation}
where $\theta$ is the contact angle, $\gamma_{\rm WH}$ is the IFT between the H$_2$O-rich and H$_2$-rich phases,
$\gamma_{\rm SH}$ is the IFT between the solid and H$_2$-rich phases,
and $\gamma_{\rm SW}$ is the IFT between the solid and H$_2$O-rich phases. 
The adhesion tensions of the H$_2$O+H$_2$+silica system were calculated based on IFTs and CAs, which are plotted in Fig. S5(a). 
Similar to the values in the H$_2$O+silica system given in Fig. S6, the adhesion tensions fall in the range from 22.8 to 46.9 mN/m. The adhesion tensions decrease significantly with temperature. For example, at 40 MPa, the adhesion tensions decrease from 45.2 to 22.8 mN/m as temperature increases from 298 to 523 K. However, the adhesion tensions change little as pressure increases. 

Fig. \ref{fig:z6} displays the density profiles of H$_2$O and H$_2$ perpendicular to the silica surface for the H$_2$O+H$_2$+silica system. Data at other pressures were given in Fig. S7.
The density profiles of H$_2$O here are similar to those reported for the H$_2$O+silica system\cite{yang2022interfacialJML1} (also see Figs. S8 (a) and (b) for results from present simulations).
The density of H$_2$O inside the droplet oscillates near the silica indicating strong adsorptions. Notably, H$_2$ also adsorbs on the silica surface inside the droplet (see Fig. \ref{fig:z6}(a)). However, the adsorption peak of H$_2$ is roughly three orders of magnitude smaller than that of H$_2$O suggesting that the interactions of the H$_2$O-silica pair are much stronger than that of the H$_2$-silica pair. In the interface between the H$_2$-rich and silica phase, density peaks of both H$_2$ and H$_2$O can be observed, while the latter is much stronger than the former (see Fig. \ref{fig:z6}(b)). 

In the interface between the H$_2$O-rich and silica phase (see  Figs. \ref{fig:z6}(a) and (c)), the magnitude of peaks and bulk densities of H$_2$O increase with decreasing temperature and increasing pressure, while those of H$_2$ increase with increasing temperature and increasing pressure.
In the interface between the H$_2$-rich and silica phase (see  Figs. \ref{fig:z6}(b) and (d)), the magnitude of peaks and bulk densities of H$_2$ increase with decreasing temperature and increasing pressure, while those of H$_2$O increase with increasing temperature and decreasing pressure. Notably, the effect of pressure on the adsorption of H$_2$O is greater at elevated temperatures. Overall, the component distributions of H$_2$O and H$_2$ change greatly with temperature, while only the profiles of H$_2$ increase greatly with pressure. Considering the relatively weak interactions between H$_2$ and silica, the above findings may explain the greater temperature effect on adhesion tensions.

We now analyze the behaviour of CA of the H$_2$O+H$_{2}$+silica based on the variations of $\gamma_{\rm WH}$ (fluid-fluid interaction) and adhesion tension (fluid-solid interaction) based on Eq. \ref{eq:LY}. For temperatures in the range from 298 to 323 K, the proportional reduction of the $\gamma_{\rm WH}$ and adhesion tension with elevated temperature explains to small temperature effect on CA. At higher temperatures, the reduction of CAs arises from the greater reduction of the $\gamma_{\rm WH}$ in contrast to that of adhesion tension. 
For temperatures in the range from 298 to 323 K, the little pressure effect on CA comes from the little changes of both $\gamma_{\rm WH}$ and adhesion tension with pressure.
At higher temperatures, the increment of CA with pressure arises more from the increase of $\gamma_{\rm WH}$, since adhesion tension changes moderately with pressure.

\subsection{Interfacial properties in the H$_2$O+H$_{2}$+kerogen system}

Our simulated CA of H$_2$O with the kerogen surface was
validated using literature data.\cite{ho2021molecular} For instance, simulations by Ho and Wang\cite{ho2021molecular} showed that the CA of H$_2$O on kerogen surface was about 42.8$\degree$ at 300 K. We obtained a similar value of 40.9$\degree$ $\pm$ 1.4$\degree$ at 298 K as shown in Fig. S4.
Fig. \ref{fig:z5}(b) shows the contact angles of H$_2$O droplet from the MD simulation for the H$_2$O+H$_{2}$+kerogen three-phase system under 298-523 K and 1-160 MPa. It can be seen that the simulated CAs fall in the range from 0.0$\degree$ to 61.9$\degree$. 
The overall behaviour of CAs in the H$_2$O+H$_{2}$+kerogen system notably differs from those in the system with silica. 
The CAs decrease with temperature, and at 523 K and 5 MPa, the kerogen surface is fully water-wet (CA = 0$\degree$).
The reduction of CA with temperature is similar to that in the H$_{2}$O+kerogen system as shown in Fig. S4.
At 298 K, as pressure increases, the CA increases first and then change little with pressure when the pressure is above around 80 MPa. At higher temperatures, the CA increases with pressure. For instance, at 523 K, the CA increases from 0$\degree$ to 31.4$\degree$ as pressure increases from 5 to 160 MPa.
The reported temperature and pressure effects on CAs here are consistent with the behaviours of CAs from MD simulations of the H$_2$O+kerogen systems in the presence of CH$_4$\cite{ho2021molecular,jagadisan2022molecular} or CO$_2$.\cite{ho2021molecular,zhou2022mechanisms}

The adhesion tensions of the H$_2$O+H$_2$+kerogen system are displayed in Fig. S5(b). 
At low pressure, the adhesion tensions are in line with the values in the H$_2$O+kerogen system given in Fig. S6.
The adhesion tensions fall in the range of 25.1 to 52.7 mN/m. 
The adhesion tensions decrease with temperature, and the reduction of adhesion tension with temperature is larger at relatively low pressures. 
Moreover, the adhesion tensions decrease with pressure, and the reduction of adhesion tension with pressure is larger at relatively low temperatures. 

Fig. \ref{fig:z7} displays the density profiles of H$_2$O and H$_2$ perpendicular to the kerogen surface for the H$_2$O+H$_2$+kerogen system. Data at other pressures were given in Fig. S9.
The density profiles of H$_2$O here are consistent with those for the H$_2$O+kerogen system as presented in Figs. S8(c) and (d).
The density of H$_2$O inside the droplet oscillates near the
kerogen. However, the H$_2$O adsorption is weaker than that on the silica surface. Furthermore, the location of the H$_2$O peak is farther from the kerogen surface than the silica surface.
The density profiles of H$_2$ inside the droplet are also shown in the insets of Figs. \ref{fig:z7}(a) and (c). The adsorption of H$_2$ is smaller than that in the silica case.
In the interface between the H$_2$-rich and kerogen phases shown in Figs. \ref{fig:z7}(b) and (d), density peaks of both H$_2$ and H$_2$O can be observed. Remarkably, here the adsorption of H$_2$O is much smaller in comparison to the corresponding data in the silica system. 
Furthermore, the effects of temperature and pressure on the density profiles in the H$_2$O+H$_2$+kerogen system are similar to those in the case with silica mentioned above.
An important difference is that the droplet tends to move away from the kerogen surface as temperature increases (see Fig. S8(c)).
It is also important to note that high pressure significantly decreases the adsorption of H$_2$O in the interface between H$_2$-rich and kerogen phases while only a moderate pressure effect is there for the silica system. This may explain the drop in adhesion tensions with pressure.

The behaviour of CAs in the H$_2$O+H$_{2}$+kerogen system can be understood as follows. 
The decrease of CA with temperature can be explained mainly by the greater reduction of $\gamma_{\rm WH}$ in contrast to a relatively small reduction of the adhesion tension, especially at high pressures. 
At low temperatures, the change of adhesion tension due to pressure increment is much larger than that of $\gamma_{\rm WH}$. Therefore, the increment of CAs at low temperatures with pressure is mainly because of the fluid-kerogen interactions. However, at high temperatures, the change of adhesion tension due to pressure increment is little while  $\gamma_{\rm WH}$ increases with pressure. Hence, the increment of CAs at high temperatures with pressure is mainly due to the fluid-fluid interactions.

Furthermore, the capillary pressure can be calculated based on the standard Young-Laplace equation, 
\begin{equation}
	P_{\rm c}= (2 \gamma_{\rm WH} {\rm cos} \theta)/r_c, 
\end{equation}
where $r_c$ is the radius of curvature at the interface.\cite{nielsen2012predicting}
The capillary forces block the upward movement of the gas stored in aquifers and their escape since the solid surfaces are hydrophilic.
For a given $r_c$, the behavior of $P_{\rm c}$ is similar to that of the adhesion tension as described above. Fig. S10 displays the calculated capillary pressures with a pore radius of 40 nm in the H$_2$O+H$_2$+silica and H$_2$O+H$_2$+kerogen system. 
The results show that the $P_{\rm c}$ falls in the range of 1.1 to 2.6 MPa.
The $P_{\rm c}$ decreases significantly with temperature. The $P_{\rm c}$ on silica changes moderately with pressure, while that on kerogen decreases with pressure and this pressure effect is greater at lower temperatures.
These data point to the fact that, for example, high temperature and pressure conditions might have the risk of H$_2$ leakage because of low capillary forces.

\section{Conclusion}

%%%%%%
%%%%%%
%%%%%%
MD simulations were carried out to provide molecular perspectives of interfacial behaviours of the H$_2$O+H$_2$ and H$_2$O+H$_2$+solid (silica or kerogen) systems over a broad range of temperatures (298 - 523 K) and pressures (1 - 160 MPa).
The combination of H$_2$ model with the INTERFACE force field and TIP4P/2005 H$_2$O model can accurately predict the interfacial tensions from the experiment.\cite{chow2018interfacial,chow2020erratum}
Interfacial properties in the H$_2$O+H$_2$ systems from MD were consistent with the predictions from the DGT coupled to the PC-SAFT EoS. The IFTs of the H$_2$O+H$_2$ system are close to the H$_2$O surface tensions and are higher than the IFTs of the H$_2$O+CO$_2$\cite{chalbaud2009interfacial}, the H$_2$O+CH$_4$\cite{wiegand1994interfacial}, and the H$_2$O+N$_2$\cite{wiegand1994interfacial} systems especially at high pressures.
In general, the calculated IFTs of the H$_2$O+H$_2$ system decrease with pressure and temperature, which is in line with the experimental data.\cite{chow2018interfacial,chow2020erratum,hosseini2022h2,slowinski1957effect,massoudi1974effect}
Interestingly, at relatively high temperatures and pressures, the IFTs increase with pressure. Furthermore, the transition pressure for the inversion of pressure effects on IFTs decreases as temperature increases.
Similar complex pressure dependences of IFTs were also reported in the H$_2$O+CO$_2$\cite{garrido2019physical}, the H$_2$O+CH$_4$\cite{wiegand1994interfacial}, and the H$_2$O+N$_2$\cite{wiegand1994interfacial} systems. 
The opposite pressure effect on IFTs can be explained by the inversion of the sign of the H$_2$ relative adsorption according to the Gibbs adsorption equation. Moreover, enrichment of H$_2$ in the interfacial regions was observed in density profiles from MD and DGT, and the enrichment is generally more pronounced at low temperatures and pressures.

The interfacial behaviours in the H$_2$O+H$_2$+silica system are noticeably different to those in the H$_2$O+H$_2$+kerogen system.
The H$_2$O CAs in the H$_2$O+H$_2$+silica and H$_2$O+H$_2$+kerogen system are in the ranges 21.8$\degree$ - 49.0$\degree$ and 0.0$\degree$ - 61.9$\degree$, respectively.
The H$_2$O CAs for the H$_2$O+H$_2$+silica and H$_2$O+H$_2$+kerogen systems increase with pressure and decrease with temperature. However, the effect of temperature and pressure on these CAs is less pronounced for the H$_2$O+H$_2$+silica system at low temperatures (298 and 373 K). 
The behaviours of CAs were understood based on the variations of IFTs in the H$_2$O+H$_2$ system (fluid-fluid interaction) and adhesion tensions (fluid-solid interaction).
The adhesion tensions decrease significantly with temperature.
The adhesion tensions on silica change moderately with pressure, while those on kerogen decrease with pressure (this pressure effect is stronger at lower temperatures).
Furthermore, the analysis of the atomic density profiles shows that the presence of H$_2$ between the H$_2$O droplet and silica/kerogen surfaces is little. 
But the adsorption of H$_2$O on the silica surface outside the H$_2$O droplet is strong, while that on the kerogen surface is relatively small.
In general, temperature and pressure have great effects on component density profiles. Nevertheless, the effect of pressure on H$_2$O adsorption on silica outside the H$_2$O droplet is small.
%%%%
This investigation may provide a fundamental understanding of the interfacial behaviours in multiphase systems containing H$_2$, which may be useful for the processes regarding the geological storage of H$_2$ in sites including depleted oil/gas reservoirs and coal seams.

%%%%%%%%%%%%%%%%%%%%%%%%%%%%%%%%%%%%%%%%%%%%%
%%%%%%%%%%%%%%%%%%%%%%%%%%%%%%%

\bigskip
\bigskip
%%%%%%%%%%%%%%%%%%%%%%%%%%%%%%%%%%%%%%%%%%%%% %%%%%%%%%%%%%%%%%%%%%%%%%%%%%%%
{\bf{ACKNOWLEDGMENTS\\[1ex]}}
The investigation is supported by the National Natural Science Foundation of China (Grant No. 42203041), the Natural Science Foundation of Jiangsu Province (Grant No. BK20221132), and the China Postdoctoral Science Foundation (Grant No. 2022M723398).
This work is also supported by the King Abdullah University of Science and Technology, Office of Sponsored Research, under Award No. OSR-2019-CRG8-4074.
%%This work is partially supported by King Abdullah University of Science and Technology (KAUST) through the grants BAS/1/1351-01, URF/1/4074-01, and URF/1/3769-01.
%The authors would like to thank anonymous reviewers for many helpful comments.

{\bf{Supporting Information\\[1ex]}}
Additional details of simulation analysis are provided in the Supporting Information.
\pagebreak
\bibliography{New}

%%%%%%%%%%%%%%%%%%%%%%%%%%%%%%%%%%%%%%%%%%%%%%%%%%%%%%%%%%%%%%%%%%%%%%%%
%%%%%%%%%%%%%%%%%%%%%  Table of Contents Graphics

\clearpage
{\renewcommand{\arraystretch}{0.7}
  \begin{threeparttable}
		\centering\caption{Charges $q$ and LJ parameters $\sigma$ and $\varepsilon$ of H$_2$O and different hydrogen models.}
		{\begin{tabular}{@{}lccc}
				\toprule
				Force or charge site & $\sigma$ (\AA) & $\varepsilon$ (kcal/mol) & $q$ (e) \\
				\bottomrule
				\textit{Water} (TIP4P/2005 model) \cite{abascal2005general} & & & \\
				O  & 3.1589 & 0.1852 & ~~0.0000   \\
				H  & 0.0000 & 0.0000  & ~~0.5564    \\
				M$_{H_2O}$  &    0.0000 & 0.0000   & $-$1.1128    \\
				& & & \\
				
				\textit{Hydrogen} (IFF model) \cite{wang2021accurate} & & & \\
				H & 2.5996 & 0.0153 & ~~0.0000  \\
				& & & \\
				%% bond length 0.7414 A 350 Kcal/mol/A^2, K(r-r_o)^2
				\textit{Hydrogen} (Hirschfelder model)  \cite{hirsehfelder1954molecular} & & & \\
				H$_2$ & 2.9700 & 0.0662 & ~~0.0000 \\
				& & & \\
				
				\textit{Hydrogen} (Alavi model)    \cite{alavi2005molecular} & & & \\
				H & 0.0000 & 0.0000 & ~~0.4932 \\
				M$_{H_2}$\tnote{*} & 3.0380 & 0.0682 & $-$0.9864 \\
				%% bond length 0.7414 A	 Rigid rigid			
				\bottomrule
		\end{tabular}}
   \begin{tablenotes}
	\small
	\item[*] The massless particle M is at the center of the H$_2$ molecule with a fixed bond length of 0.7414 \AA.
\end{tablenotes}
\end{threeparttable}

\clearpage
\begin{figure}[tb]
	\begin{centering}
		\includegraphics[width=1.1\textwidth]{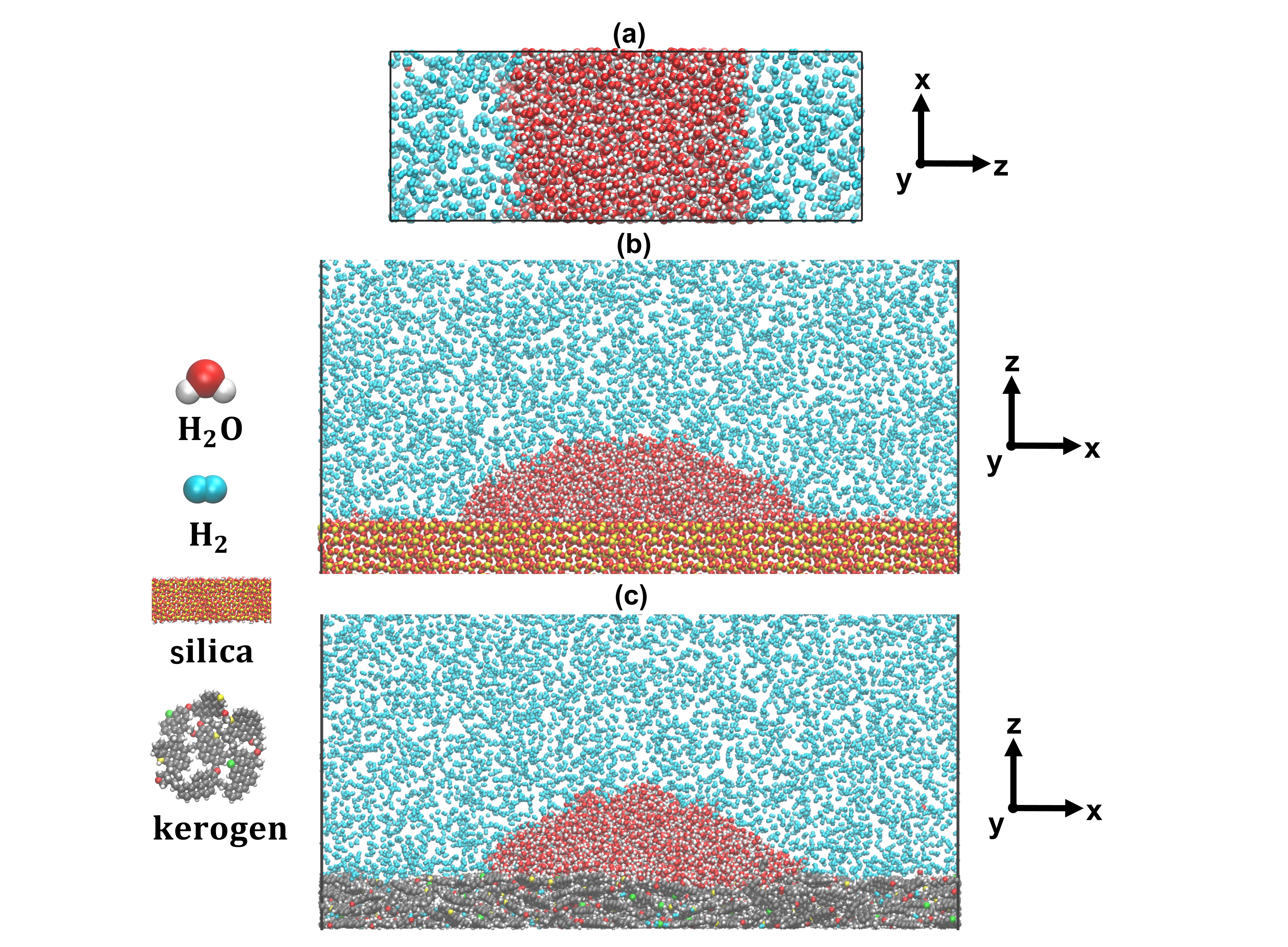}
		\caption{Equilibrium snapshots of the (a) H$_2$O+H$_2$ system, (b) H$_2$O+H$_2$+silica system, and (c) H$_2$O+H$_2$+kerogen system at 373 K and 80 MPa.
	    Color code for H$_2$O: O, red; H, white.
		Color code for H$_2$: H, cyan. 
		Color code for silica: Si, yellow; O, red; H, white. 
		Color code for kerogen: C, grey; O, red; N, yellow; S, green; H, white. 
		}
		\label{fig:z1_vmd}
	\end{centering}
\end{figure}

\clearpage
\begin{figure}[tb]
	\begin{centering}
		\includegraphics[width=0.6\textwidth]{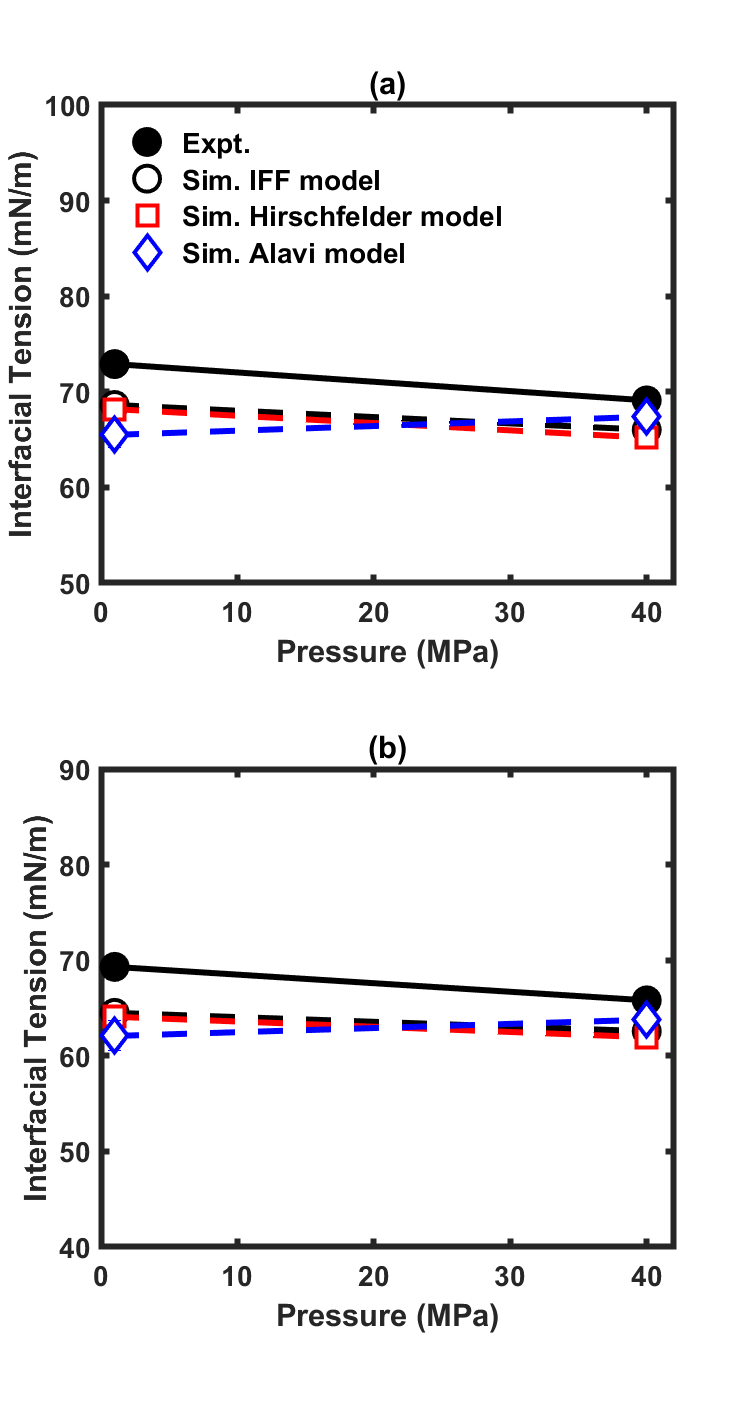}
		\caption{IFTs as a function of pressure in the H$_2$O+H$_2$ system at (a) 298 and (b) 323 K. The data from simulation with IFF\cite{wang2021accurate}, Hirschfelder\cite{hirsehfelder1954molecular}, and Alavi\cite{alavi2005molecular} H$_2$ models are shown as open circle, square, and diamond symbols, respectively. The experimental data from Chow et al.\cite{chow2018interfacial,chow2020erratum} are shown as solid symbols. The lines are a guide for the eye. Error bars smaller than the symbol size are not displayed.
		}
		\label{fig:z2}
	\end{centering}
\end{figure}

\clearpage
\begin{figure}[tb]
	\begin{centering}
		\includegraphics[width=0.6\textwidth]{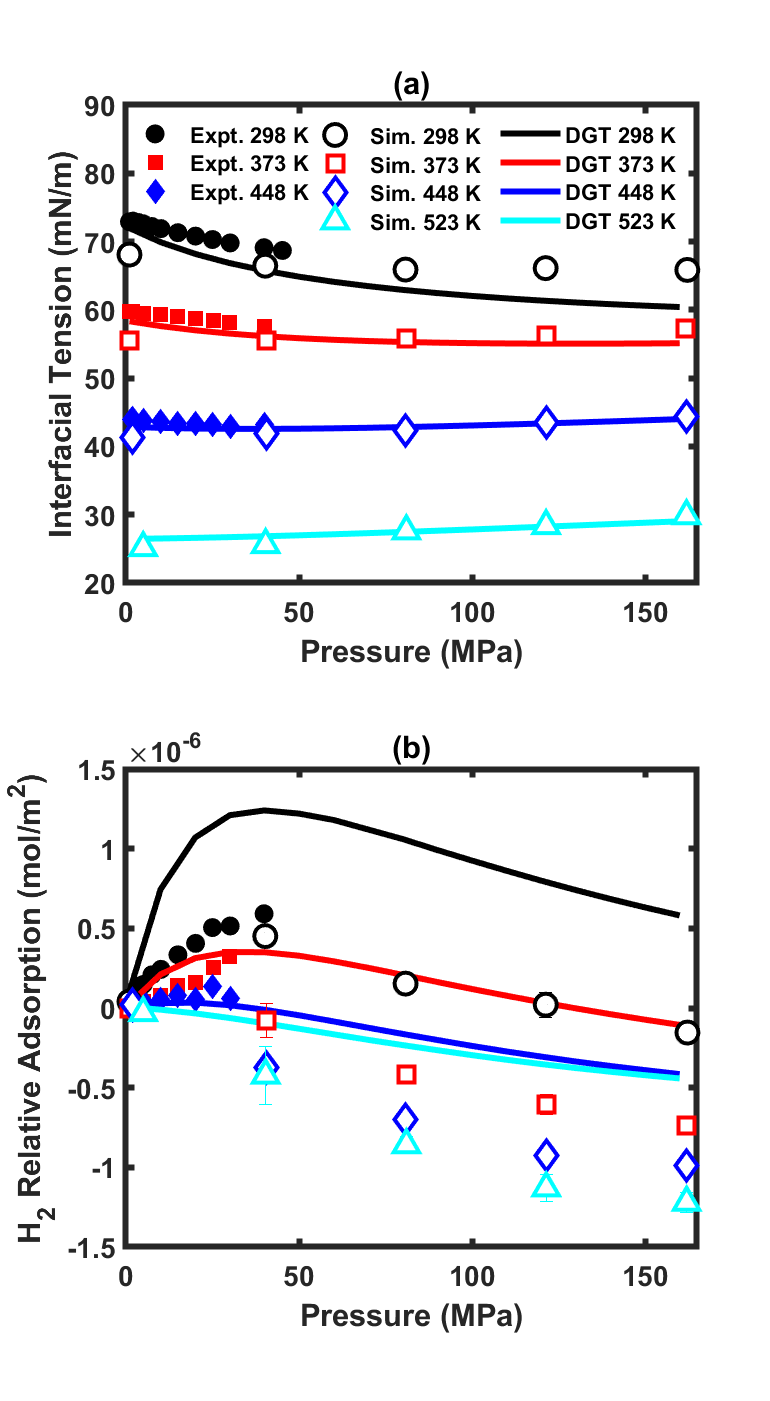}
		\caption{(a) Pressure dependence of the IFT for the H$_2$O+H$_2$ system at various temperatures. (b) Pressure dependence of the relative adsorption of H$_2$ at various temperatures.
		The open symbols denote the data from the MD simulations and the estimates from DFT with the PC-SAFT EoS are shown as lines. 
		The experimental data from Chow et al.\cite{chow2018interfacial,chow2020erratum} are shown as solid symbols.
		Error bars smaller than the symbol size are not displayed.}
		\label{fig:z3}		
	\end{centering}
\end{figure}

\clearpage
\begin{figure}[tb]
	\begin{centering}
		\includegraphics[width=1.0\textwidth]{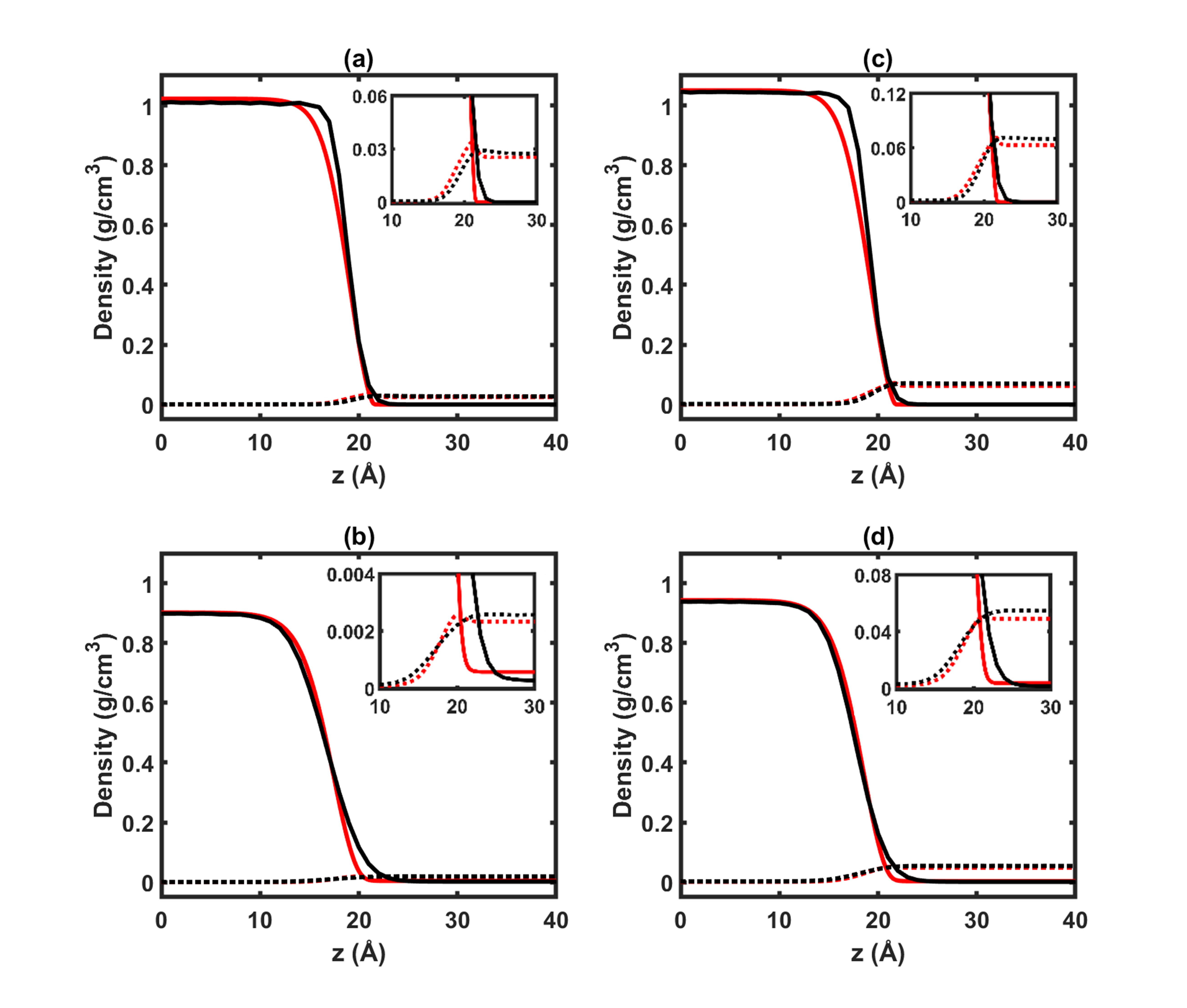}
		\caption{Equilibrium distributions of H$_2$O (solid lines) and H$_2$ (dotted lines) in the H$_2$O+H$_2$ system at (a) 298 K and 40 MPa, (b) 448 K and 40 MPa, (c) 298 K and 160 MPa, and (d) 448 K and 160 MPa. 
		The black and red colors denote MD and DGT data, respectively.
		}
		\label{fig:z4}
	\end{centering}
\end{figure}

\newpage
%%%%%%%%%%%%%%%%%%%% 
\begin{figure}[tb]
	\begin{centering}
		\includegraphics[width=0.5\textwidth]{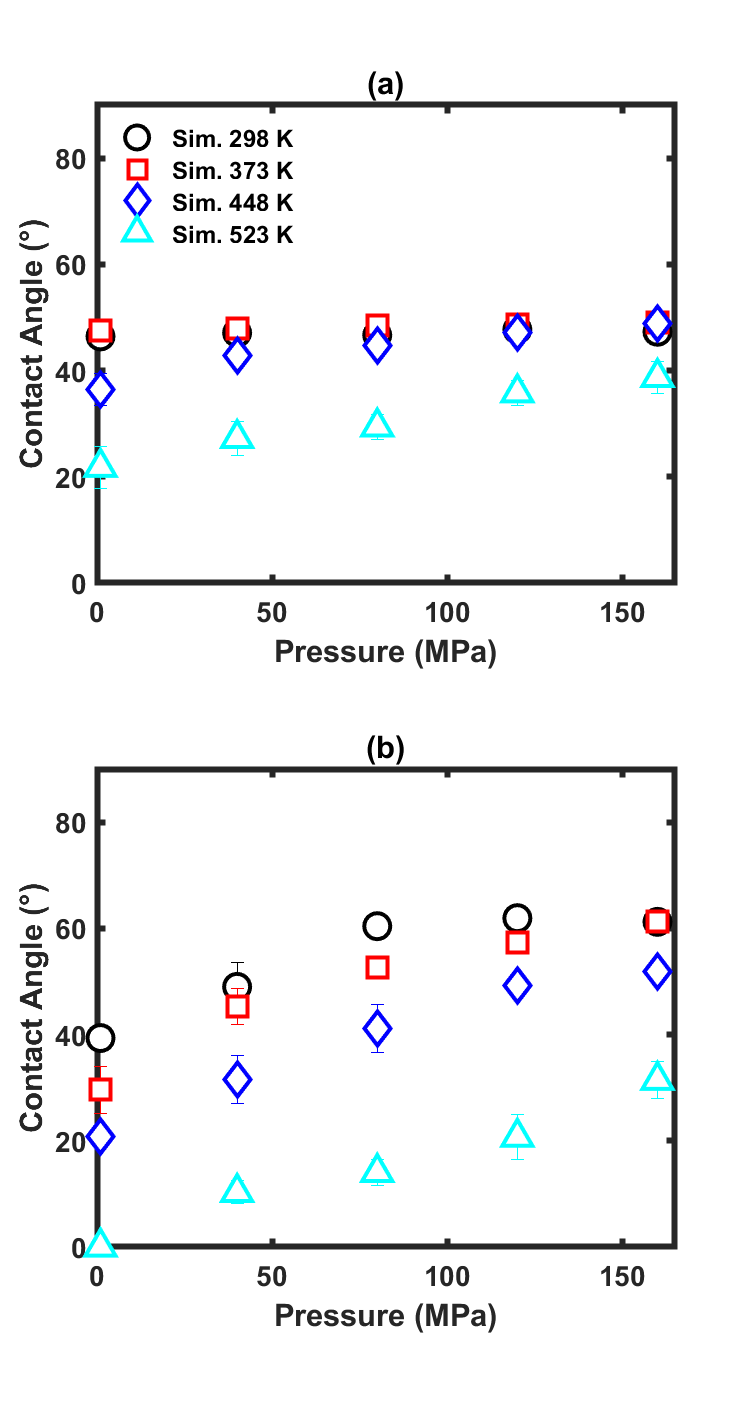}
		\caption{Contact angles from MD simulations as a function of pressure in the (a) H$_2$O+H$_2$+silica system and (b) H$_2$O+H$_2$+kerogen system at various temperatures. Error bars smaller than the symbol size are not displayed.
		}
			\label{fig:z5}
	\end{centering}
\end{figure}

\clearpage
\begin{figure}[tb]
	\begin{centering}
		\includegraphics[width=1.1\textwidth]{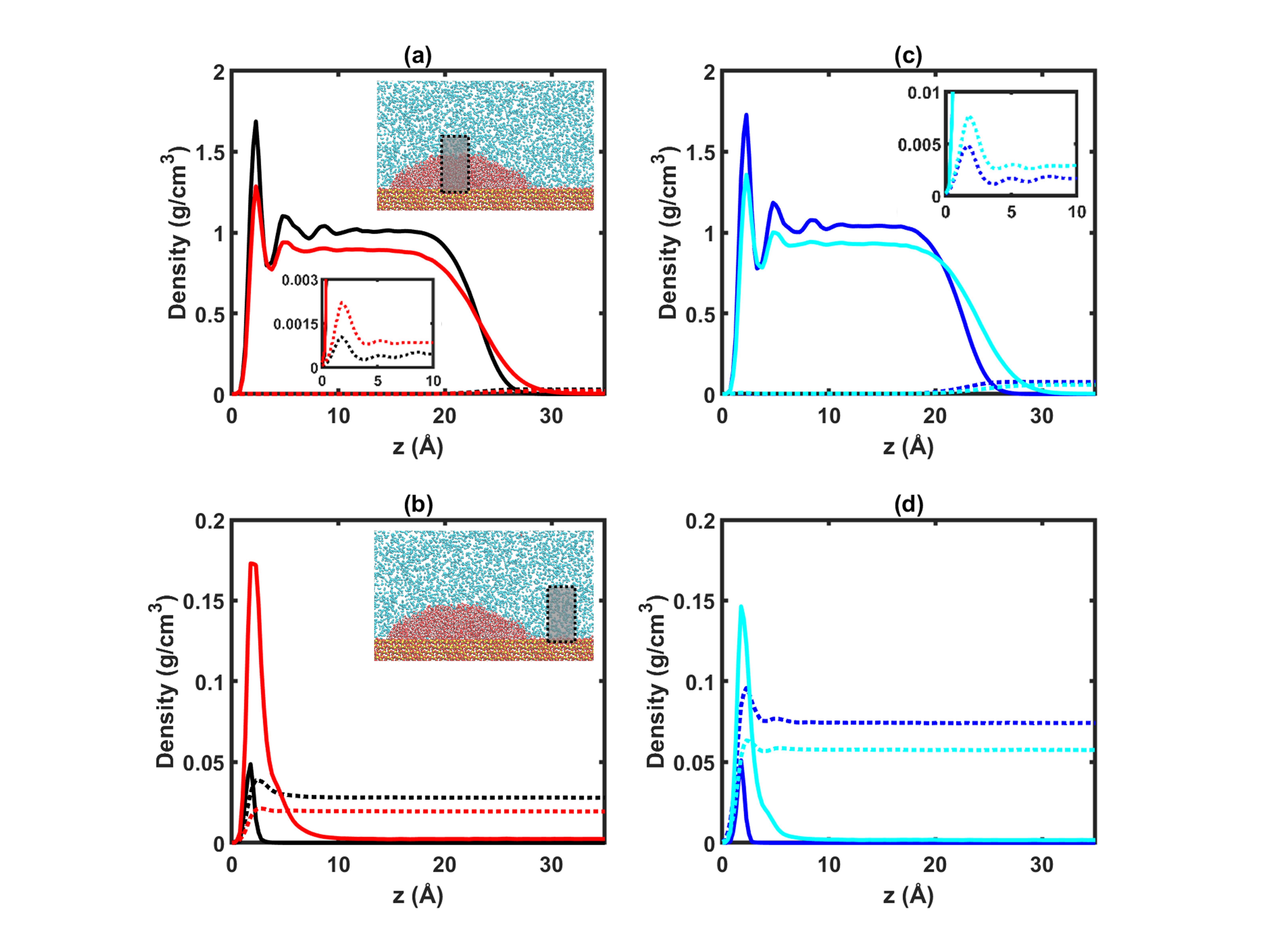}
		\caption{Equilibrium distributions of H$_2$O (solid lines) and H$_2$ (dotted lines) normal to the silica surface from MD simulations of the H$_2$O+H$_2$+silica system: 298 K, 40 MPa (black); 448 K, 40 MPa (red); 298 K, 160 MPa (blue); and 448 K, 160 MPa (cyan). 
		The density profiles in the region of the H$_2$O droplet (shaded region of the inset snapshot in (a)) are shown in (a) and (c), while those in the H$_2$-rich region (shaded region of the inset snapshot in (b)) are shown in (b) and (d).
	}
		\label{fig:z6}
	\end{centering}
\end{figure}

\clearpage
\begin{figure}[tb]
	\begin{centering}
		\includegraphics[width=1.1\textwidth]{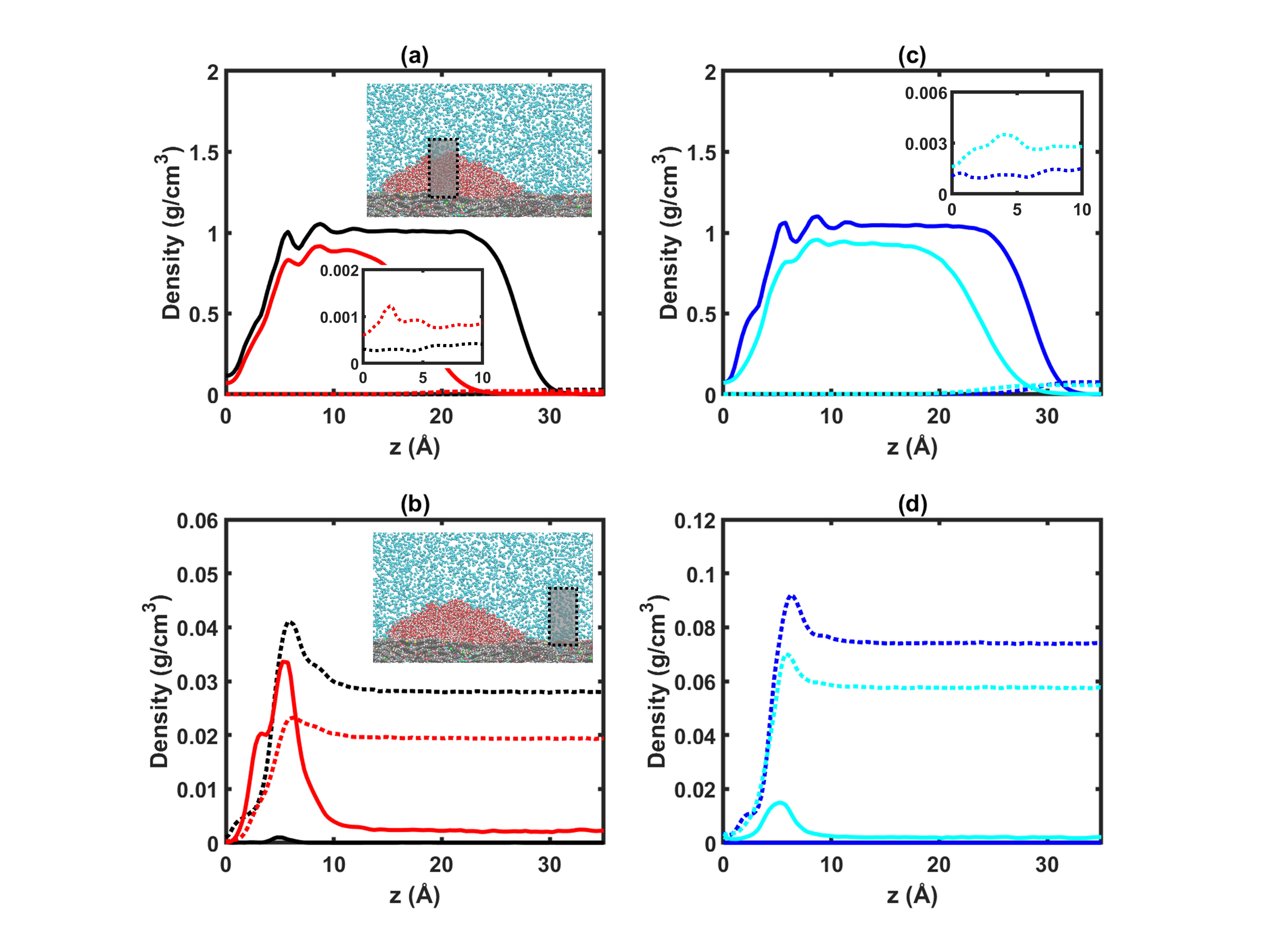}
		\caption{Equilibrium distributions of H$_2$O (solid lines) and H$_2$ (dotted lines) normal to the kerogen surface from MD simulations of the H$_2$O+H$_2$+kerogen system: 298 K, 40 MPa (black); 448 K, 40 MPa (red); 298 K, 160 MPa (blue); and 448 K, 160 MPa (cyan). 
		The density profiles in the region of the H$_2$O droplet (shaded region of the inset snapshot in (a)) are shown in (a) and (c), while those in the H$_2$-rich region (shaded region of the inset snapshot in (b)) are shown in (b) and (d).
		}
		\label{fig:z7}
	\end{centering}
\end{figure}

\end{document}